\documentclass[preprint,prb,floats,amsfonts,showpacs,eqsecnum,superscriptaddress,floatfix,multicol aps,showpacs]{revtex4}
\usepackage{amssymb}
\usepackage{epsfig}
\usepackage{amsmath}
\usepackage{psfrag}
\usepackage{color}
\usepackage{graphicx}
\usepackage{subfigure}
\usepackage{bm}

\usepackage{epsfig}

\newcommand{\be}{\begin{equation}}
\newcommand{\ee}{\end{equation}}
\newcommand{\bea}{\begin{eqnarray}}
\newcommand{\eea}{\end{eqnarray}}

\begin{document}
\draft
\title{The Fermi liquid nature of the ground state of double quantum dots in parallel from a $1/N$ expansion}

\author{Manas Kulkarni$^{1,2}$ and Robert Konik}

\affiliation{Department of Condensed Matter Physics and Material Science,
Brookhaven National Laboratory, Upton, NY-11973\\
$^2$Department of Physics and Astronomy, Stony Brook University, Stony Brook, NY 11794-3800}
\begin{abstract}
A large-N diagrammatic approach is used to study coupled quantum dots
in a parallel geometry. We show that the Friedel sum rule (FSR) holds at lowest
order in a 1/N expansion for this system, thereby suggesting
that the ground state is a Fermi liquid.  Using the FSR together with
the dot occupancy, we compute the dot system's conductance.
Our finding that the $1/N$ expansion indicates the system is a Fermi liquid
is in agreement with both prior results based on the Bethe ansatz
and slave boson mean field theory.
\end{abstract}

\maketitle

\section{Introduction}

Strong correlations in impurity problems have been of tremendous theoretical
interest.\cite{glazman-review,lee-2010,sela-2009,sela-2009qcp,Ireneusz-2010,malecki-sela-affleck,oguri,short,long,zitko,
zitko-bonea-2006,hofstetter,ding,ding-ye,le-hur,vernek,ding-kim,luis-2006,luis-2008,sela-affleck-2009p,komnik}  
This interest has been spurred by the ability to engineer
semiconducting quantum dots which are highly tunable.\cite{gold,cronenwett,gold1,gold_nano,
chen-2004,chen-review,sigrist-2004,holleitner-2004,hubel-2008}  In such realizations
of quantum dots, both the gate voltages of the quantum dot as well as the tunneling
amplitudes between the dots and the leads can be tuned.  This gives one the ability
to study these systems throughout their entire parameter space.  While the first
reported dot systems involved single dots,\cite{gold,cronenwett} more complicated dot structures can
now be fabricated.\cite{gold_nano, gold1,chen-2004,chen-review,sigrist-2004,holleitner-2004,hubel-2008}

With this ability to engineer multi-quantum dot systems comes the ability to
realize more exotic forms of Kondo physics.  In double dot systems, for example, one can explore
the competition between the Kondo effect and the Ruderman-Kittel-Kasuya-Yosida
(RKKY) interaction.\cite{lee-2010,sela-2009,sela-2009qcp,Ireneusz-2010,malecki-sela-affleck,oguri,short,long,zitko,
zitko-bonea-2006,hofstetter,ding,ding-ye,le-hur,vernek,ding-kim,luis-2006,luis-2008,sela-affleck-2009p,komnik,kulk}
Because of the non-perturbative nature of the Kondo effect at low
temperatures, one might expect this competition to be non-trivial.  We argue that
it is in fact so for double dots arranged in parallel.

The geometry that we wish to consider is sketched in Fig. 1.  Here two closely spaced, single level dots
do not directly interact with one another, either via tunneling or via a capacitive coupling.
They are however connected to one another via hopping processes involving the two leads.
We will want to consider a situation where a total of two electrons sit on the two dots and
where the leads are symmetrically coupled to the two dots.  In such a scenario only one
continuum channel of electrons (the even combination of the two leads) couples to the two dots.
In such a system the Kondo effect and the RKKY interaction are antithetical to one another.
The Kondo effect will promote many-body singlet formation while the RKKY interaction will tend
to encourage direct triplet formation between the two electrons on the two dots.

\begin{figure}
\begin{center}
\includegraphics[scale=0.6,angle=-90]{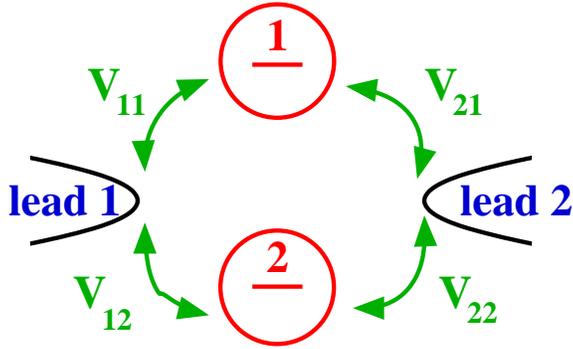}
\caption{A schematic of closely spaced quantum dots
arranged in parallel. The bare energy level on the dots are $\tilde\epsilon_{d_{1/2}}$.}
\end{center}
\end{figure}

\begin{figure}
\begin{center}
\includegraphics[scale=0.4,angle=0]{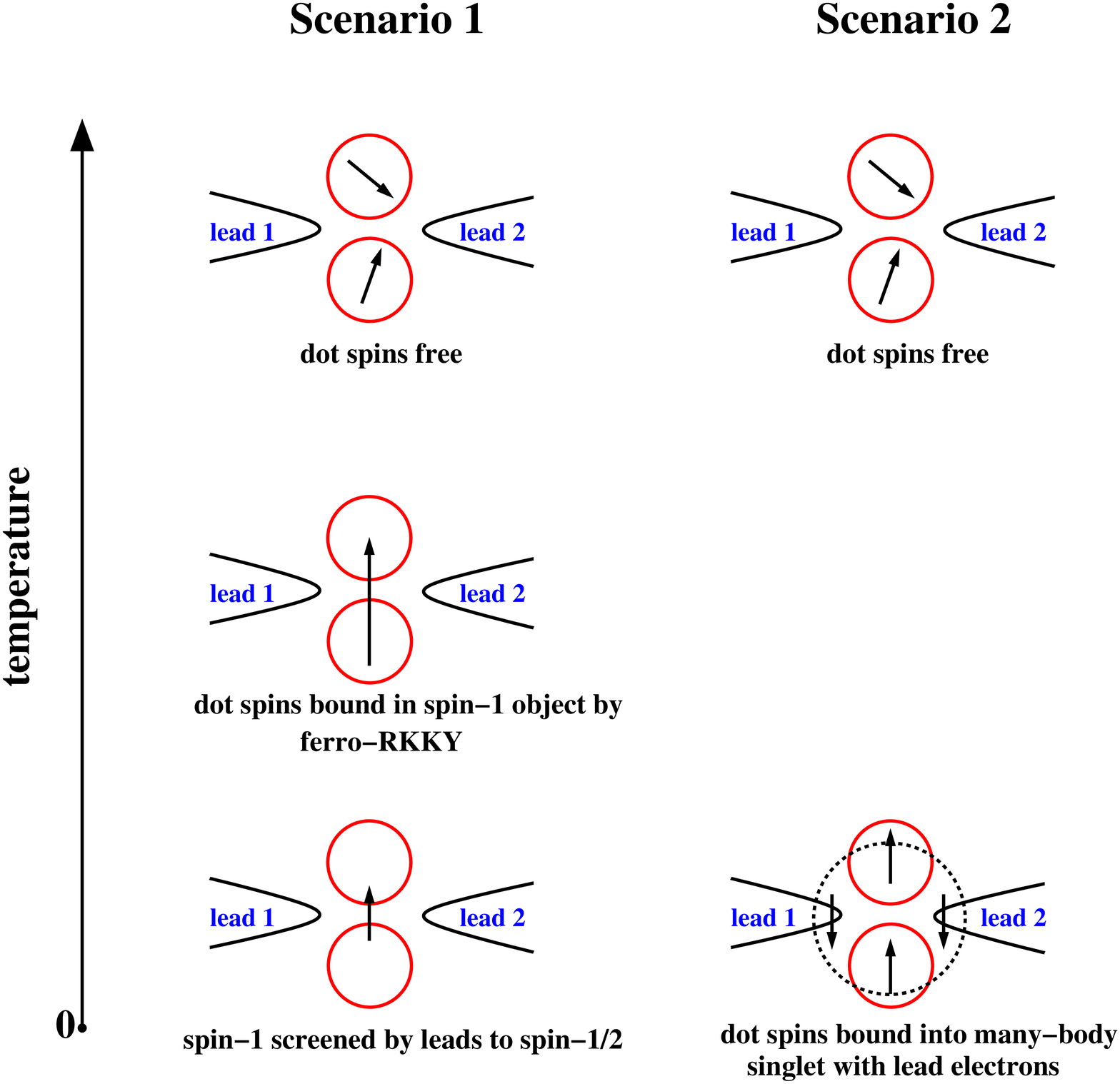}
\caption{Two possible scenarios for how the ground state forms
in doubly occupied double dots in parallel as the temperature is lowered.  In Scenario 1 the
two electrons first form a spin-1 object via a ferromagnetic RKKY interaction which is then
screened to spin-1/2 by the one effective channel of the leads.  In Scenario 2 
the RKKY interaction does not dominate the Kondo effect and the two electrons on
the dots form a many-body singlet with the lead electrons.
}
\end{center}
\end{figure}

One can then imagine (at least) two possible scenarios for ground state formation as one lowers the temperature
(see Fig. 2).
In the first scenario the RKKY interaction dominates and at some intermediate temperature
binds the two electrons on the two dots into
a triplet.  As the temperature is furthered lowered the Kondo effect then comes into play partially screening the
effective spin-1 object on the dots down to an effective spin-1/2 (i.e. an underscreened Kondo effect).
Here the physics is non-Fermi liquid with low temperature quantities such as conductance having a logarithmic
dependence in temperature.
This scenario has been argued to hold in a number of studies including those based on a numerical
renormalization group (NRG).\cite{zitko, zitko-bonea-2006, ding, ding-ye}
The particular implementation of the NRG there, however, has been criticized.\cite{kulk}

A second scenario can be envisioned characterized by a lack of a clear separation of the RKKY effect from
Kondo physics (i.e. the persistence of higher temperature RKKY-induced triplet formation down
to temperatures below a putative Kondo temperature).  In this scenario the low temperature state
is an overall many-body singlet.  At low temperatures any effective binding by an RKKY interaction into
a triplet state comes undone.  Rather the two electrons on the two dots participate in the singlet
by screening one another in conjunction with the lead electrons.  The physics in this scenario is sharply differentiated
from the first by being Fermi liquid.  In particular the deviations in the conductance with temperature go
as $T^2$.  This scenario has been supported by both Bethe ansatz computations\cite{short,long} as well as slave boson mean field
theory treatments.\cite{kulk}

It is the purpose of this paper to provide supporting evidence for the second scenario.  In it we present
a $U=\infty$, $1/N$ treatment of the problem. The method of $1/N$ expansion is 
well known\cite{bickers,brandt} and has been used in the context of quantum dots/impurities.\cite{aditi,pcoleman,pcoleman1} 
This treatment is distinct from the use of slave boson mean field theory (SBMFT) to argue the validity of the second scenario.  
While SBMFT is itself a $1/N$ technique, it represents some (uncontrolled) summation of $1/N$ diagrams.  Here instead
we consider a $1/N$ expansion in a systematic fashion.  In particular we argue that this $1/N$ expansion
is consistent with the Friedel sum rule -- a hallmark of Fermi liquid physics and so indicative of the
second scenario.  (In SBMFT, the Friedel
sum rule is automatically satisfied by virtue of the quadratic form of the action.)

The paper is organized as follows. In Section II
we briefly describe the double-dot model we are interested in studying and
the $1/N$ method used. In Section III
we present results for the dot Green's functions, the system's partition
function, and the dot occupancy.  In Section IV we show that the Friedel sum rule holds and so
the system appears to be a Fermi liquid,
the central result of this paper.  In Section V we compute the conductance using the Friedel sum
rule and compare it to slave boson mean field. We then draw conclusions in Section IV.
Details related to computing diagrams in the $1/N$ expansion are relegated to appendices.

\section{Model of double dots in parallel}

The Anderson like Hamiltonian that we study is given by

\begin{eqnarray}\label{eIIi}
H &=& -i\sum_{l\sigma}\int_{-\infty}^{+\infty}dxc_{l\sigma}^{\dagger}\partial_{x}c_{l\sigma}+
\sum_{l\sigma\alpha}V_{l\alpha}\left(c_{l\sigma\alpha}^{\dagger}d_{\sigma\alpha}+{\rm h.c.}\right)\cr\cr
&& \hskip .5in + \sum_{\sigma\alpha}\epsilon_{d\alpha}n_{\sigma\alpha}
 + \sum_{\alpha}U_{\alpha}n_{\uparrow\alpha}n_{\downarrow\alpha}.
\end{eqnarray}
The $c_{l\sigma}$ specify electrons with spin $\sigma$ living on
the two leads, $l=L,R$. The $d_{\alpha\sigma}$ specify electrons
found on the two dots $\alpha=1,2$. Electrons can hop from the leads
to dots with tunneling strength $V_{l\alpha}$.
We suppose
that there is no interdot Coulomb repulsion and that tunneling between
the two dots is negligible.
We further consider this model in the limit
that the strength of the
Coulomb repulsion on the two dots, $U_{\alpha}$, is taken to $\infty$ thus precluding
any double occupancy.
The constraint of no double-occupancy is fulfilled
by first introducing the slave boson formalism, one slave
boson for each dot:
\begin{equation}\label{eIIii}
d_{\sigma\alpha}=b_{\alpha}^{\dagger}f_{\sigma\alpha}.
\end{equation}
Here $f_{\sigma\alpha}$ is the pseudofermion (denoted later in Feynman diagrams by dashed lines) which annihilates
one ``occupied state'' on dot $\alpha$ and $b_{\alpha}^{\dagger}$
is a bosonic operator (denoted later by wavy lines) which creates an empty state
on dot $\alpha$.
We subject these new degrees of freedom to the constraint,
\begin{equation}\label{eIIiii}
Q_\alpha \equiv b^\dagger_\alpha b_\alpha + \sum_\sigma f^\dagger_\sigma f_\sigma,
\end{equation}
which we enforce by adding two Lagrange multipliers $\lambda_{1}$ and $\lambda_{2}$ to the Hamiltonian
\begin{eqnarray}\label{eIIii}
\delta H = i\lambda_1 Q_1 + i\lambda_2 Q_2.
\end{eqnarray}
We thus understand the partition function of the system to be computed via
\begin{equation}
Z = \int^{\pi T}_{-\pi T} \frac{\beta}{2\pi}d\lambda_i e^{i\beta(\lambda_1+\lambda_2)}{\rm Tr}e^{-\beta(H+i\sum_i\lambda_iQ_i)}.
\end{equation}
In treating quantities in this model, we will perform a systematic diagrammatic expansion in $1/N$ where
$N$ is the degeneracy of each dot.

We suppose the dot-lead couplings, $V_{l\alpha}$, satisfy the following ratio condition,
\begin{equation}
\frac{V_{1\alpha}}{V_{2\alpha}}=\frac{V_{1\alpha^{\prime}}}{V_{2\alpha^{\prime}}}=\lambda.
\end{equation}
This condition ensures that only one effective channel of lead electrons
couples to the two dots.  To see this we reexpress the Hamiltonian in terms of even and odd channels,
$$
c_{e/o}=(V_{1/2,\alpha}c_1 \pm V_{2/1,\alpha}c_2)/\sqrt{2\Gamma_\alpha},
$$ 
where $\Gamma_\alpha = (V^2_{1\alpha}+V^2_{2\alpha})/2$.
In terms of these new degrees of freedom, the Hamiltonian divides itself into an even and an odd sector,
\begin{eqnarray}\label{diageIIiii}
H &=& {\cal H}_e + {\cal H}_o;\cr\cr
{\cal H}_e &=& -i\sum_{l\sigma}\int^\infty_{-\infty} dx ~c^\dagger_{e\sigma}\partial_x c_{e\sigma}
+ \sum_{\sigma\alpha}\sqrt{2\Gamma_{\alpha}}(c^\dagger_{e\sigma\alpha}d_{\sigma\alpha} + {\rm h.c}) \cr\cr
&& + \sum_{\sigma\alpha}\epsilon_{d\alpha}n_{\sigma\alpha} +
\sum_{\alpha} U_{\alpha} n_{\uparrow\alpha}n_{\downarrow\alpha};\cr
{\cal H}_o &=& -i\sum_{l\sigma}\int^\infty_{-\infty} dx ~c^\dagger_{o\sigma}\partial_x c_{o\sigma},
\end{eqnarray}
and we see that the odd sector of the Hamiltonian is decoupled from the dots.

\section{Summary of Computations}

In this section we will compute the Green's functions of the dot degrees of freedom, the dot partition function,
and the dot occupancy to leading
order and where necessary to prove the FSR the subleading order in $1/N$.

\subsection{Dot Greens functions}

In terms of the slave bosons and fermions, the dot temperature Green's functions are defined as
\begin{equation}\label{eIIIi}
G_{\alpha,\beta}=\frac
{\langle b_{\alpha}^{\dagger}(\tau)f_{\sigma\alpha}(\tau)f_{\sigma\beta}^{\dagger}(0)b_{\beta}(0)e^{-\intop_{0}^{\beta}H_{V}(\tau)d\tau}\rangle_{0}}
{\langle e^{-\intop_{0}^{\beta}H_{V}(\tau)d\tau}\rangle _{0}},
\end{equation}
where the $\alpha,\beta$ indicate which dot the electron is sitting on and $H_V$ is the part of the Hamiltonian coupling the leads
to the dots.
As computed in the appendices, the lowest order contributions in $1/N$ to these Green functions are
\begin{eqnarray}\label{eq:g11O1}
G_{11}^{O(1)}(i\omega_{n}) &=& \frac{K_1(E_{01})}{i\omega_{n}-T_{A_{1}}};\cr\cr
G_{12}^{0(\frac{1}{N})}(i\omega_{n}) &=& \frac{K_{1}(E_{01})}{i\omega_{n}-T_{A_{1}}}\left[-i\sqrt{\Gamma_{1}\Gamma_{2}}\right]\frac{K_{2}(E_{02})}{i\omega_{n}-T_{A_{2}}},
\end{eqnarray}
where
\begin{eqnarray}
\Sigma_{1,2}(z) &=& 2N\Gamma_{1,2}\sum_{k}\frac{f_{k}}{z+\epsilon_{k}-\epsilon_{d_{1,2}}}\label{eq:sigma12}\cr\cr
&=& \frac{N\Gamma_{1,2}}{\pi}\log\bigg|\frac{\epsilon_{d_{1,2}}-z}{D}\bigg| - i\theta(z-\epsilon_{d_{1,2}})N\Gamma_{1,2};\cr\cr
K_{1,2}(E_{01,02}) &=& \left[1-\frac{\partial\Sigma_{1,2}}{\partial z}\right]^{-1}\mid_{z=E_{01,02}}.
\end{eqnarray}
In Eqn. (\ref{eq:g11O1}) one can make the replacement
$1\leftrightarrow2$ to get the corresponding expressions for $G_{22}^{O(1)}(i\omega_{n})$
and $G_{21}^{O(\frac{1}{N})}(i\omega_{n})$.  We see that the interdot correlation function, $G_{12}$,
first sees a contribution at ${\cal O}(1/N)$.
Here $\epsilon_{k}$ denotes the energy of the conduction electrons and
$f_{k}=\frac{1}{e^{\beta k}+1}$ is the Fermi function.
Finally, we note that $E_{01,2}$ is the most negative solution
of $\omega-Re\Sigma_{1,2}(\omega)=0$ and we correspondingly define $T_{A_{1,2}}$ as
$T_{A_{1,2}}=\epsilon_{d_{1,2}}-E_{01,2}$.

For the purposes of proving the Friedel sum rule, we will need not only these contributions to the dot Green's
functions but the imaginary part of the subleading (${\cal O}(1/N)$) 
correction to $G_{11}$ and $G_{22}$ as well.  These are as follows:
\begin{equation}\label{eq:g11im}
{\rm Im}\left[G_{11}^{O(1/N)}\right]=-\frac{K_{1}^2(E_{01})\Gamma_{1}}{\left(\omega-T_{A_{1}}\right)^{2}}-\pi\theta(-\omega-T_{A_{1}})R(\omega)-\pi\theta(\omega-T_{A_{1}})\left[S(\omega)+T(\omega)\right],
\end{equation}
where $S(\omega)\,,R(\omega)$, and $T(\omega)$ are defined by
\begin{eqnarray}
R(\omega) &=& \frac{K_{1}(E_{01})}{N}\left[\frac{N\Gamma_{1}}{\pi}\right]^{2}\int_{T_{A_{1}}}^{-\omega}\frac{dk}{\left(\omega+\epsilon_{k}-T_{A_{1}}\right)^{2}}\frac{1}{\left(\epsilon_{k}-\left(\frac{N\Gamma_{1}}{\pi}\right)\log(\frac{\epsilon_{k}}{T_{A_{1}}}-1)\right)^{2}+\left(N\Gamma_{1}\right)^{2}};\cr\cr
S(\omega) &=& \frac{K_{1}(E_{01})}{N}\left[\frac{N\Gamma_{1}}{\pi}\right]^{2}\frac{1}{\left(\omega-T_{A_{1}}\right)^{2}}\int_{T_{A_{1}}}^{\omega}\frac{dk}{\left(\epsilon_{k}-\left(\frac{N\Gamma_{1}}{\pi}\right)\log(\frac{\epsilon_{k}}{T_{A_{1}}}-1)\right)^{2}+\left(N\Gamma_{1}\right)^{2}};\cr\cr
T(\omega) &=& \frac{K_{1}(E_{01})}{N}\left[\frac{N\Gamma_{1}}{\pi}\right]^{2}\left[\frac{1}{T_{A_{1}}}-\frac{1}{\omega}\right]\frac{1}{\left[\omega-T_{A_{1}}+\left(\frac{N\Gamma_{1}}{\pi}\right)\log(\frac{\omega}{T_{A_{1}}})\right]^{2}}.
\end{eqnarray}
We obtain $G_{22}^{O(1/N)}$ by swapping $1\leftrightarrow2$.

The above expression for ${\rm Im}G_{11}^{O(1/N)}$ undergoes a dramatic simplification when $\omega$ is set to 0.  We find
\begin{equation}
{\rm Im}\left[G_{11}^{O(1/N)}(\omega=0)\right]=-\frac{K_1^2(E_{01})\Gamma_1}{T_{A_1}^2}.
\label{FSRlimit}
\end{equation}
This simplification will be useful for proving the Friedel sum rule.

\subsection{Dot occupancy}
The partition function and the dot occupancy obey the simple relation
\begin{equation}
n_{d1,2}=-\frac{1}{\beta}\frac{\partial \log {\cal Z}}{\partial \epsilon_{d1,2}}.
\end{equation}
The partition function at ${\cal O}(1)$ is given by
\begin{equation}
{\cal Z}=e^{-\beta E_{01}} e^{-\beta E_{02}},
\end{equation}
and $E_{01/02}$ are defined as in Section III A.  After some algebra we then find that
\begin{equation}
n_{d1,2}=\frac{\mu_{1,2}}{1+\mu_{1,2}},
\label{nd1nd2}
\end{equation}
where
\begin{equation}
\mu_{1,2}=\frac{N \Gamma_{1,2}}{\pi T_{A1,2}}.
\end{equation}
This $O(1)$ computation of the dot occupancy is sufficient for proving the Friedel sum rule at leading order in $1/N$.

\section{Proof of the Friedel sum rule for double quantum dots}

The Friedel sum rule (FSR) states for a system with N-fold degeneracy that the scattering phase, $\delta(E)$,
of an electron at the Fermi surface $E=0$ is given in terms of the total occupancy, $n_d$, of the dots by
\begin{equation}
\delta(0)=\frac{\pi n_{d}}{N}.\label{eq:FSR}
\end{equation}
The Friedel sum rule implies that the zero temperature linear response conductance is then given by
\begin{equation}
G=2\frac{e^{2}}{h}\frac{4\lambda^2}{(1+\lambda^2)^{2}}\sin^2(\frac{\pi n_d}{N}),\label{eq:linear}
\end{equation}
where $\lambda = V_{1\alpha}/V_{2\alpha}$, $\alpha=1,2$.

From Ref. \onlinecite{mwprl}, the linear response conductance can be expressed in terms of the dot Green's functions as
follows:
\begin{equation}
G=-2\frac{e^{2}}{h}\frac{2}{(1+\lambda^2)}\mbox{Im}[\mbox{Tr}\left\{ \Gamma_{L}G^{r}\right\} ]
\label{eq:meir}
\end{equation}
where $G^{r}$ is the full retarded Green's function matrix in dot-space
and $\Gamma_{L}$ is the bare hybridization matrix.  These are defined as
\begin{equation}
G^{r}=\left[\begin{array}{cc}
G_{11}^{r} & G_{12}^{r}\\
G_{21}^{r} & G_{22}^{r}\end{array}\right]\label{eq:G_matrix},
\end{equation}
and
\begin{equation}
\Gamma_{L}=\left[\begin{array}{cc}
V_{11}^{2} & V_{11}V_{12}\\
V_{11}V_{12} & V_{12}^{2}\end{array}\right]\label{eq:Gamma_M}.
\end{equation}
Thus to show that the FSR holds we need to show that
\begin{equation}
2\frac{e^{2}}{h}\frac{4\lambda^2}{(1+\lambda^2)^{2}}\sin^{2}\left(\frac{\pi n_{d}}{N}\right)=-2\frac{e^{2}}{h}\frac{2}{(1+\lambda^2)}\mbox{Im}[\mbox{Tr}\left\{ \Gamma_{L}G^{r}\right\} ]\label{eq:toprove}\end{equation}
holds in terms of a $1/N$ expansion.

The left hand side of (\ref{eq:toprove}) to $O(\frac{1}{N^{2}})$ is equal to
\begin{eqnarray}
\frac{4\lambda^2}{(1+\lambda^2)^{2}}\sin^{2}\left(\frac{\pi n_{d}}{N}\right)&=&
\frac{4\lambda^2}{(1+\lambda^2)^{2}}\sin^{2}\left(\frac{\pi\left(n_{d1}+n_{d2}\right)}{N}\right)\nonumber \\
&=&\frac{4\lambda^2}{(1+\lambda^2)^{2}}\frac{\pi^{2}}{N^{2}}\biggl\{n_{d1}^{2^{\mbox{O(1)}}}+n_{d2}^{2^{\mbox{O(1)}}}
+ 2n_{d1}^{\mbox{O(1)}}n_{d2}^{\mbox{O(1)}}\biggr\},
\label{eq:2sin}
\end{eqnarray}
where we only consider the dot occupancies to ${\cal O}(1)$.
The right hand side of (\ref{eq:toprove}) to the same order
equals
\begin{eqnarray}
-\frac{2}{(1+\lambda^2)}\mbox{Im}[\mbox{Tr}\left\{ \Gamma_{L}G^{r}\right\} ]&=&-\frac{2}{(1+\lambda^2)}\mbox{Im}\biggl\{ V_{11}^{2}G_{11}^{\mbox{O(1/N)}}+V_{11}V_{12}G_{21}^{\mbox{O(1/N)}}\nonumber\\&+&V_{11}V_{12}G_{12}^{\mbox{O(1/N)}}+
V_{12}^{2}G_{22}^{\mbox{O(1/N)}}\biggr\}
\label{eq:2trc}
\end{eqnarray}
where all the Green's functions were evaluated to the given order in the previous section.  Note that
${\rm Im}G_{11/22}^{\mbox{O(1)}}=0$ and so does not appear in the above.
Using Eqs. \ref{FSRlimit} and \ref{nd1nd2} one can then readily verify
that the FSR (\ref{eq:toprove}) holds for double dots to leading order in a ${\cal O}(1/N)$ expansion.
This then matches the proof of the FSR in single dots in a large-N expansion performed in Ref. \onlinecite{bickers}.

\section{Evaluation of conductance using FSR and $1/N$ diagrams}

\begin{figure}
\includegraphics[scale=0.5]{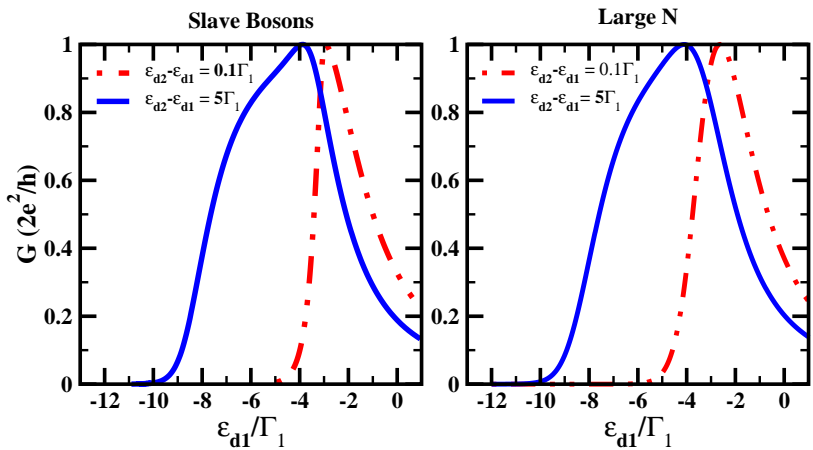}
\caption{\label{fig:condp}Plot showing the linear response conductance
for two different values of separation between the dots' chemical potential.
In the left panel we present results arrived at from the application of SBMFT
done in Ref. \onlinecite{kulk} and in the right panel the results coming from the large-N
expansion.}
\end{figure}

Having proven the FSR, we will now use it to compute the linear conductance, $G$, i.e. we will use
the expression
\begin{equation}
G=2\frac{e^{2}}{h}\frac{4\lambda^2}{(1+\lambda^2)^{2}}\sin^{2}
\left(\frac{\pi\left[n_{d_{1}}+n_{d_{2}}\right]}{N}\right),\label{eq:linear_cond_diag}
\end{equation}
where $n_{d_{1}}$ and $n_{d_{2}}$ are calculated from diagrams of
the partition function $Z$ (see Section III B).

In Fig. 3 we present the conductance for two different spacings between the chemical
potentials of the two dots (one large and one small).  We compare the large-N computation done in this paper with
the results of Ref. \onlinecite{kulk}.  We find remarkable agreement between the two approaches.
Not only are the general trends in the conductances similar, but more detailed features such
as inflection points are reproduced.  As the dot chemical potentials are lowered we see that
the conductance vanishes.

In Fig. 4 we plot the corresponding total dot occupancies both from the $1/N$ expansion
(to lowest order in $1/N$) and SBMFT (from Ref. \onlinecite{kulk}).  We again see
that there is a close correspondence between the two methods.  This is notable
as in computing the dot occupancies from the $1/N$ expansion we only keep ${\cal O}(1)$
diagrams while the SBMFT accounts for diagrams at all orders of $1/N$.

\begin{figure}
\includegraphics[scale=0.5]{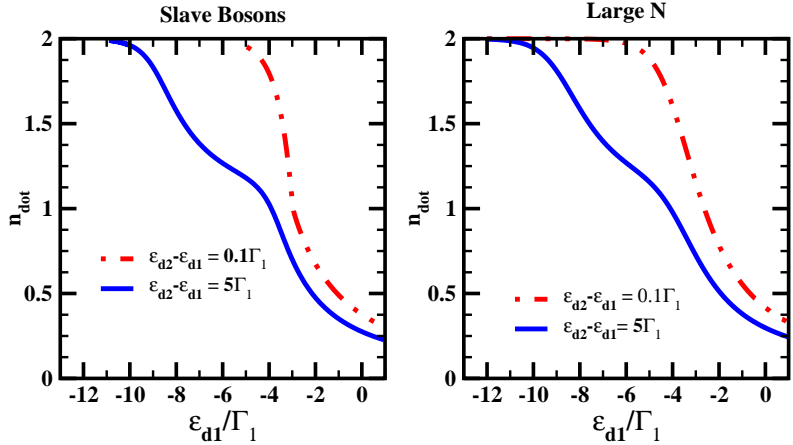}
\caption{\label{fig:condp}Plot showing the total occupation of the dots
for two different values of separation between the dots' chemical potential.
As with Fig. 2 we show the values of the dot occupancy both computed in SBMFT
and in the 1/N expansion. }
\end{figure}

\section{Discussions and Conclusions}

In this paper we have studied a double-impurity model using
a $1/N$ expansion where $N$ is the degeneracy of each
dot level. Using this $1/N$ expansion we demonstrated that the Friedel sum rule holds, so providing complementary
evidence that double dots in parallel in the absence of interdot and capacitive coupling
have Fermi liquid ground states.  In previous works we have argued that
this ground state is Fermi liquid using both the Bethe ansatz\cite{short,long} and SBMFT.\cite{kulk}
This then provides a complementary piece of evidence for this Fermi liquid nature.

Such evidence is needed because there exists in the literature an opposite set of results coming
from numerical renormalization group analyses.\cite{zitko,zitko-bonea-2006,ding,ding-ye}
In Ref. \onlinecite{kulk} we have suggested a possible reason for this discrepancy, namely in discarding
modes that arise from the use of logarithmic basis, the NRG discards additional effective screening channels
that would completely screen any putative spin-1 object that might form under a ferromagnetic RKKY interaction.
But as also pointed out in Ref. \onlinecite{kulk}, another possible reason for the discrepancy exists.  In treating
the double dots in parallel all of these methodologies (Bethe ansatz, SBMFT, $1/N$, NRG) treat the dots as
point scatterers in a continuum media.  Then with two dots in close proximity to one another there exists the
possibility that the physics might be influenced by the particular implementation of a UV cutoff that the methodology
employs.  If the physics was sensitive to this implementation, one would expect to find non-universal results.
It is notable then that we were able to conclude in this study that the $1/N$ expansion produces results (i.e.
a Fermi liquid ground state) equivalent to that of SBMFT and the Bethe ansatz.  This finding
then gives additional weight to the universal nature of the Fermi liquid ground state of double dots in parallel.

We will end on a tangential note by stressing the remarkable similarity between results obtained for a double
dot in a (low order) $1/N$ expansion and SBMFT.  This would not be necessarily expected as the SBMFT represents
some uncontrolled sum over diagrams at all orders in $1/N$ and $N$ is merely 2.  It is an interesting
question why these higher order
diagrams are suppressed strongly even when $N=2$.  We note though that while the $1/N$ expansion agrees quantitatively
with SBMFT and while both of these methods agree on a qualitative level with
the Bethe ansatz, these methods
do not produce fine details of the conductance of the double dots brought out by the Bethe ansatz (see Ref. \onlinecite{kulk}),
presumably the most trustworthy method of the three as it represents an exact solution of the problem.

Acknowledgements: M.K. was supported by the NSF under Grant No. DMR- 0906866.
R.M.K. acknowledges support by the US DOE under Contract No. DE-AC02-98 CH 10886.
We thank A. Tsvelik and P. Coleman for useful discussions.

\newpage

\appendix

\section{ $O(1)$  diagrams of  $G_{11}(i\omega_{n})$}
\label{app:g11low}
\begin{figure}
\includegraphics[scale=.6]{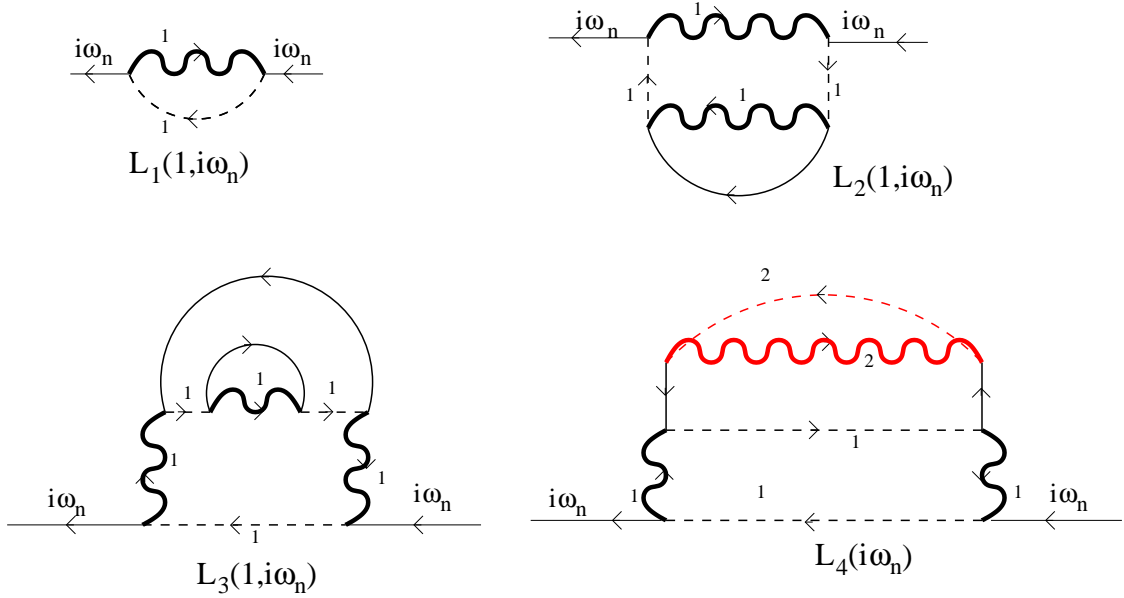}
\caption{\label{L1L2}Given above are the diagrams involved in the evaluation of the numerator of the expression for $G_{11}$ at
both ${\cal O}(1)$ ($L_1(1,i\omega_n)$) and ${\cal O}(1/N)$ ($L_2(1,i\omega_n), L_3(1,i\omega_n)$, and $L_4(i\omega_n)$).  
The argument $j$ of $L_{1,2,3}(j,i\omega_n)$ indicates the diagram involves only dot electrons of species j.
(And so $L_4$ does not have this argument as it involves both.) 
The wavy thick lines indicate dressed boson propagators, the dashed lines, bare pseudofermion lines, and the solid (internal) lines,
conduction electron propagators. The labels and colors ((1,black) and (2,red)) indicate the lines belong to
the bosons, $b$, and fermions, $f$, of different dots.}
\end{figure}

\begin{figure}
\includegraphics[scale=.6]{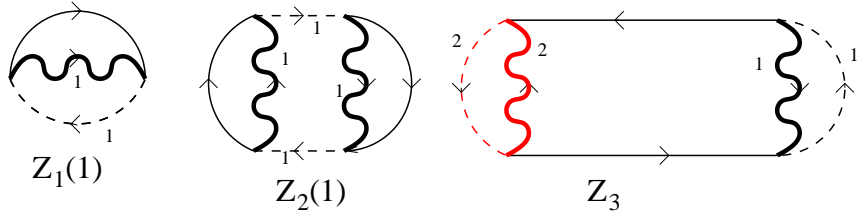}
\caption{\label{noleg}Given above are the diagrams involved in the evaluation of the denominator of the expression for $G_{11}$ at
both ${\cal O}(1)$ ($Z_1(1)$) and ${\cal O}(1/N)$ ($Z_2(1)$ and $Z_3$).}
\end{figure}

Following the definition of the Green's function in Eq. \ref{eIIIi}, $G_{11}$, can be written as
\begin{equation}\label{eAi}
G_{11}^{O(1)}(i\omega_{n})=\frac{L_{1}(1,i\omega_{n})Z_1(1)}{Z_{1}(1)Z_{1}(2)},
\end{equation}
a product of diagrams.
These diagrams
are given in Figs. \ref{L1L2}  and \ref{noleg} and 
these are the only diagrams at $O(1)$ that survive the projection
$$
\int^{\pi T}_{-\pi T} \frac{\beta}{2\pi}d\lambda_1 d\lambda_2 e^{i\beta\sum_i{\lambda_i}}
$$
involved in computing any correlation function.
The argument $j=1,2$ of $Z_1(j)$ and $L_1(j,i\omega_n)$ indicates the diagram involves dot electrons
on dot $j$.
$L_{1}(1,i\omega_n)$ involves a dressed boson propagator, $B(z)$, for dot 1 with self energy
$\Sigma_{1}$ -- defined in Eq. \ref{eq:sigma12}.  This dressing occurs already at ${\cal O}(1)$
and is straightforwardly obtained via a Dyson equation (see Fig. 4
in Ref. \onlinecite{bickers} as well as Ref. \onlinecite{brandt}):
$$
B(z) = (B_0(z)^{-1}-\Sigma_{1,2})^{-1},
$$
where $B_{0}(z)= 1/(z-i\lambda)$ is the bare propagator of the boson line.
$L_1(1,i\omega_n)$ is then equal to
\begin{eqnarray}\label{eAii}
L_{1}(1,i\omega_n)=\frac{1}{\beta}\sum_{\nu_{m}}\int_{-\pi T}^{\pi T}\frac{\beta d\lambda_1}{2\pi}e^{i\beta\lambda_1}\frac{1}{i\nu_m-i\lambda_1-\Sigma_{1}(i\nu_{m}-i\lambda_1)}\frac{1}{i(\omega_{n}+
\nu_{m})-i\lambda_1-\epsilon_{d1}}.
\label{app1}
\end{eqnarray}
In Eqn. \ref{eAii}
the summation over Matsubara boson frequencies can be performed
by the usual technique of converting the summation into a contour integration.  Doing so we find
\begin{eqnarray}\label{eAiii}
L_1(1,i\omega_n)=\int_{-\pi T}^{\pi T}\frac{\beta d\lambda_1}{2\pi}e^{i\beta\lambda_1}b(E_{01}+i\lambda_1)\frac{K_{1}(E_{01})}{i\omega_{n}-T_{A_{1}}}.
\end{eqnarray}
Here $E_{01}$ is the most negative solution of $z-Re\Sigma_{1}(z)=0$, $b(z)=(e^{\beta z}-1)^{-1}$ is
the Bose function, and $K_1$ is given by
\begin{eqnarray}\label{eAiv}
K_1(E_{01}) & = & [1-\frac{\partial\Sigma_{1}}{\partial z}]^{-1}\mid_{z=E_{01}}.
\end{eqnarray}
In order to perform the $\lambda$ integration we note that the Bose
function can be expanded
as $b(z+i\lambda_{1,2})=e^{-\beta(z+i\lambda_{1,2})}+O(e^{-2i\beta\lambda_{1,2}})$.
From this it follows straightforwardly that
\begin{equation}\label{eAv}
L_{1}(1,i\omega_{n})=\frac{e^{-\beta E_{01}}K_{1}(E_{01})}{i\omega_{n}-T_{A_{1}}}.
\end{equation}
To determine $E_{01}$, we note that
\begin{equation}\label{eAvi}
\Sigma_{1}(z)=2N\Gamma_{1}\sum_{k}\frac{f_{k}}{z+\epsilon_{k}-\epsilon_{d_{1}}}
\end{equation}
evaluates to
\begin{eqnarray}\label{eAvii}
{\rm Re}\Sigma_{1}(\omega) &=& \frac{N\Gamma_{1}}{\pi}\log\left|\frac{\epsilon_{d_{1}}-\omega}{D}\right|;\cr\cr
{\rm Im}\Sigma_{1}(\omega) &=& -\theta(\omega-\epsilon_{d1})N\Gamma_1.
\end{eqnarray}
Eq. \ref{eAvii} is obtained after converting the summation into an integration over k which brings in a factor of $\frac{1}{2\pi}$. As $E_{01}$ is the most negative solution of $\omega-{\rm Re}\Sigma_{1}(\omega)=0$
which satisfies $E_{01}<\epsilon_{d_1}$ we can write
$T_{A_{1}}=\epsilon_{d_{1}}-E_{01}$
and see that $T_{A_{1}}$ is always positive. $T_{A_{1}}$ is, in effect, the renormalized dot chemical
potential of dot 1.

We will now compute $Z_{1}(1)$ a bubble diagram (see Fig. \ref{L1L2}).
It differs from $L_1(1)$ in that it also involves a conduction electron line (represented
as a solid line in Fig. \ref{L1L2}) and involves no external frequency. It is important to note that the boson propagator in
this case is different from the boson propagator in the diagram represented
by $L_1(1,i\omega_n)$. The boson propagator in $Z_1(1)$ involves summing over a class of diagrams,
each of which comes with an additional factor of $1/n$.\cite{bickers,brandt}
In order to make this sum correspond to the usual
geometric series\cite{bickers,brandt} we introduce a \textit{coupling constant parameter} $g$
by the transformation of the coupling constant $\Gamma_{1}\rightarrow g^{2}\Gamma_{1}$. By suitably differentiating the series
by $g$ one can turn this series into a regular geometric series.\cite{mahen,fetter,orland}
We thus arrive at the expression:
\begin{eqnarray}\label{eAviii}
Z_1(1)&=&-2\int_{-\frac{\pi}{\beta}}^{+\frac{\pi}{\beta}}\frac{\beta d\lambda_1}{2\pi}e^{i\beta \lambda_1}
\intop_{0}^{1}\frac{dg}{g}\sum_{\nu_{m},\omega_{n},k}\cr\cr
&&\hskip -.4in \times\frac{2Ng^{2}\Gamma_{1}}{\beta^{2}}
\frac{1}{i\nu_{m}-i\lambda_2-\Sigma_{1}(i\nu_{m}-i\lambda_1,g)}
\frac{1}{i(\nu_{m}+\omega_{n})-(\epsilon_{d2}+i\lambda_1)}\frac{1}{i\omega_{n}-\epsilon_{k}}.
\end{eqnarray}
After performing the Matsubara summation over $\omega_{n}$ and the projection integral, and transforming
the expression for the Matsubara sum over $\nu_{m}$ to a contour integral, we obtain
\begin{equation}\label{eAix}
Z_{1}(1)=-\beta\intop_{o}^{1}\frac{dg}{g}\intop_{\Gamma_{2}}\frac{dz}{\pi i} e^{-\beta z}
\frac{\Sigma_{1}(z,g)}{z-\Sigma_{1}(z,g)}.
\end{equation}
Proceeding with the coupling constant integration in Eqn. \ref{eAviii}, we find,
\begin{equation}\label{eAx}
Z_{1}(1)=\beta\intop_{\Gamma_{2}}\frac{dz}{2\pi i} e^{-\beta z}
\log(\frac{z-\Sigma_{1}(z)}{z}).
\end{equation}
If we now perform an integration by parts
and then compute the contour integration about the
pole at $z=E_{02}$, we find,
\begin{equation}\label{eAxi}
Z_1(1)=e^{-\beta E_{01}}.
\end{equation}
We finish up then with
\begin{equation}\label{eAxiii}
G_{11}^{O(1)}(i\omega_{n})=\frac{K_{1}(E_{01})}{i\omega_{n}-T_{A_{1}}}.
\end{equation}
A similar expression for $G_{22}^{O(1)}(i\omega_{n})$ is found by interchanging all
species indices $1\leftrightarrow 2$.

\section{\label{sec:g11next}$O(1/N)$ contribution to $G_{11}(i\omega_n)$}
$G_{11}^{O(1/N)}$ is given by the following set of contributions,
\begin{eqnarray}
G_{1,1}^{O(1/N)} &=& \frac{(L_2(1,i\omega_n)+L_3(1,i\omega_n))Z_1(2)+L_1(1,i\omega_n)Z_2(2)+L_4(i\omega_n)}{Z_1(1)Z_1(2)}\cr\cr
&& - \frac{L_1(1,i\omega_n)Z_1(2)(Z_1(1)Z_2(2)+Z_1(2)Z_2(1)+Z_3)}{Z_1(1)^2Z_1(2)^2}\cr\cr
&& \hskip -.5in = \frac{L_2(1,i\omega_n)+L_3(1,i\omega_n)}{Z_1(1)}+\frac{L_4(i\omega_n)}{Z_1(1)Z_1(2)}
- \frac{L_1(1,i\omega_n)Z_2(1)}{Z_1(1)^2} - \frac{L_1(1,i\omega_n)Z_3}{Z_1(1)^2Z_1(2)}.
\end{eqnarray}
The first and third terms form the ${\cal O}(1/N)$ contribution to $G_{11}$ of a single dot while the second and fourth terms reflect
the presence of a second dot.  We consider the latter first.

\subsection{Evaluation of $L_4(i\omega_n)$ and $Z_3$}
We will first consider the diagrams which contain both dot electron species, $L_4(i\omega_{n})$
and $Z_3$ (see Figs. 5 and 6). We will show that both of them go to zero in the low temperature limit.
Let us look at $L_4(i\omega_{n})$ first:
\begin{eqnarray}
L_4(i\omega_n) &=& \frac{4N\Gamma_1\Gamma_2}{\beta^3}\sum_{\nu_m,\nu_m',\omega_n',k,k'}
\frac{1}{i(\nu_{m}+\omega_{n})-(\epsilon_{d_{1}}+i\lambda_{1})}\frac{1}{(i\nu_{m}-i\lambda_{1}-\Sigma_{1}(i\nu_{m}-i\lambda_{1}))^{2}} \cr\cr
&& \times \frac{1}{i\omega_{n}^{\prime}-\epsilon_{k}}\frac{1}{i\omega_{n}^{\prime}-\epsilon_{k^{\prime}}}
\frac{1}{i(\omega_{n}^{\prime}+\nu_{m})-(\epsilon_{d_{1}}+i\lambda_{1})} \cr\cr
&& \times \frac{1}{i(\omega_{n}^{\prime}+\nu_{m}^{\prime})-(\epsilon_{d_{2}}+i\lambda_{2})}
\frac{1}{i\nu_{m}^{\prime}-i\lambda_{2}-\Sigma_{2}(i\nu_{m}^{\prime}-i\lambda_{2})}.
\label{eq:L6_first}
\end{eqnarray}
Performing the Matsubara sums over $\nu'_m$ and $\omega_n'$, we arrive at an expression given by (dropping terms
that will disappear once we apply the projection):
\begin{eqnarray}\label{eCiii}
L_4(i\omega_{n}) & = & -2\frac{4N\Gamma_{1}\Gamma_{2}}{\beta} 
\sum_{\nu_{m},k,k^{\prime}}\frac{1}{i(\nu_{m}+\omega_{n})-(\epsilon_{d_{1}}+i\lambda_{1})}\cr
 && \hskip -1in \times\frac{1}{(i\nu_{m}-i\lambda_{1}-\Sigma_{1}(i\nu_{m}-i\lambda_{1}))^{2}}\frac{f(\epsilon_k)}{\epsilon_k-\epsilon_{k'}}
\frac{1}{i\nu_m+\epsilon_k-(i\lambda_1+\epsilon_{d1})}\cr\cr
&& \times \bigg[ \frac{K(E_{02})e^{-\beta(i\lambda_2+E_{02})}}{\epsilon_k-T_{A2}}+
\frac{e^{-\beta(\epsilon_{d2}-\epsilon_k+i\lambda_2)}}{\epsilon_{d2}-\epsilon_k-\Sigma_2(-\epsilon_k+\epsilon_{d2})}+\cdots\bigg].
\end{eqnarray}
Of the two terms in the square brackets in the above expression, the first dominates in the low temperature
limit as it is proportional to $e^{-\beta E_{02}}$ and $E_{02}<\epsilon_{d2}-\epsilon_k$.  We can thus neglect the second
term in this limit.  There are other such terms which are subdominant as $T$ goes to 0 and which are indicated by
the ellipses.

If we perform the final Matsubara sum together with the $k'$ sum we obtain an expression involving terms of the form
\begin{equation}
\int dk f(\epsilon_k) \log\frac{D+\epsilon_k}{D-\epsilon_k}\frac{1}{(\epsilon_k-T_{A2})(\epsilon_k-T_{A1})^2},
\end{equation}
and 
\begin{equation}\label{eqn}
\intop_{-D}^{0}\log [\frac{D+\epsilon_{k}}{D-\epsilon_{k}}]
\frac{1}{(\epsilon_{k}-T_{A_{2}})}\frac{1}{(\epsilon_{k}-T_{A_{1}})}.
\end{equation}
It is straightforward to show that these integrals vanish as $\log(D)/D$ in the large bandwidth limit.

Now we will calculate the bubble diagram $Z_3$ (see Fig. 6).
We have
\begin{eqnarray*}
Z_3 & \propto & \sum_{\nu_{m},\nu_{m}^{\prime},\omega_{n},k,k^{\prime}}\frac{1}{i\omega_{n}-\epsilon_{k}}\frac{1}{i\omega_{n}-\epsilon_{k^{\prime}}}
 \frac{1}{i\nu_{m}-i\lambda_{1}-\Sigma_{1}(i\nu_{m}-i\lambda_{1})}\cr\cr
&& \times \frac{1}{i\nu_{m^{\prime}}-i\lambda_{2}-\Sigma_{2}(i\nu_{m^{\prime}}-i\lambda_{2})}
\frac{1}{i(\nu_{m}+\omega_{n})-(\epsilon_{d_{1}}+i\lambda_{1})}\frac{1}{i(\omega_{n}+\nu_{m}^{\prime})-(\epsilon_{d_{2}}+i\lambda_{2})}.
\end{eqnarray*}
Performing the Matsubara sums as well as the $k'$ sum leaves us with an expression again involving terms of the form
found in Eqn. (\ref{eqn}) and so like $L_4(i\omega_n)$, $Z_3$ goes to zero in the infinite bandwidth $D$ limit.

\subsection{Evaluation of remaining diagrams contributing to $G_{11}^{O(1/N)}$}

In the zero temperature, large bandwidth limit, $G_{11}^{O(1/N)}$ reduces to 
\begin{equation}
G_{11}^{O(1/N)} = \frac{L_2(1,i\omega_n)+L_3(1,i\omega_n)}{Z_1(1)}- \frac{L_1(1,i\omega_n)Z_2(1)}{Z_1(1)^2}.
\end{equation}
But this is precisely the $O(1/N)$ correction that is seen in single level dots.  We can thus use Ref. \onlinecite{bickers}
to complete the evaluation of $G_{11}^{O(1/N)}$:
\begin{eqnarray}\label{eCxix}
G_{1,1}^{O(1/N)}(i\omega_{n})&=&Q^{\prime}(i\omega_{n})+B_{1}(i\omega_{n})+B_{2}(i\omega_{n})\cr\cr
Q^{\prime}(i\omega_{n})
&=&\frac{1}{i\omega_{n}-T_{A_{1}}}\frac{\partial}{\partial(z)}[\Sigma_{1}^{1/N}(z)(\frac{z-E_{01}}{z-\Sigma_{1}(z)})^{2}]_{z=E_{01}};\cr
B_{1}(i\omega_{n}) &=&
\frac{K_{1}(E_{01})}{(i\omega_{n}-T_{A_{1}})^{2}}\Biggl[2\Gamma_{1}\sum_{k}
\frac{1-f(k)}{\left(E_{01}+i\omega_{n}-\epsilon_{k}-\Sigma_{1}(E_{01}+i\omega_{n}-\epsilon_{k})\right)}\cr
&&\hskip 1in -K_{1}(E_{01})\Sigma_{1}^{1/N}(E_{01})]\Biggr];\cr
B_{2}(i\omega_{n}) & = & 2\Gamma_{1}K_{1}(E_{01})\bigg(
-\sum_{k}\frac{f(k)}{(\epsilon_{k}-T_{A})^{2}\left(E_{01}-i\omega_{n}+\epsilon_{k}-\Sigma_{1}(E_{01}-i\omega_{n}+\epsilon_{k})\right)}
\cr
&&\hskip .9in + 2N\Gamma_{1}\sum_{k,k^{\prime}}f_{k}(1-f_{k^{\prime}})
\frac{1}{(\epsilon_{k}-T_{A_{1}})^{2}}\frac{1}{(i\omega_{n}+\epsilon_{k}-\epsilon_{k^{\prime}}-T_{A_{1}})}\cr
&&\hskip 1.2in \times\frac{1}{(E_{01}+\epsilon_{k}-\epsilon_{k^{\prime}}-\Sigma_{1}(E_{01}+\epsilon_{k}-\epsilon_{k^{\prime}}))^{2}}\bigg).
\end{eqnarray}
We have used the same notation as in Ref. \onlinecite{bickers}.

\section{\label{sec:g12low} $O(1/N)$ diagrams for $G_{12}(i\omega_{n})$ }
\begin{figure}
\includegraphics{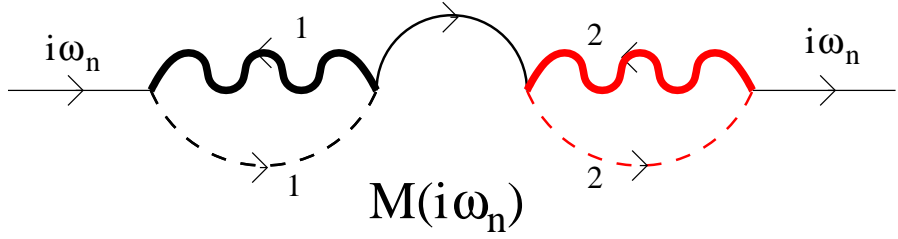}
\caption{\label{fig:M1}Diagram $M(i\omega_{n})$ contributing to $G_{12}(i\omega_n)$ at $O(1/N)$.}
\end{figure}

The lowest order contribution to
$G_{12}(i\omega_{n})$ occurs at $O(1/N)$ and can be written as
\begin{equation}
G_{12}^{O(1/N)}(i\omega_{n})=\frac{M(i\omega_{n})}{Z_1(1)Z_1(2)}.\label{eBi}
\end{equation}
We already have $Z_1(1/2)$ from Appendix A. The
diagram representing $M(i\omega_{n})$ is given in Fig. \ref{fig:M1}.
Using calculations similar to Appendix \ref{app:g11low} one easily arrives at
\begin{equation}
M(i\omega_{n})=\frac{e^{-\beta E_{01}}K_{1}(E_{01})}{i\omega_{n}-T_{A_{1}}}(-i\sqrt{\Gamma_1 \Gamma_2})\frac{e^{-\beta E_{02}}K_{2}(E_{02})}{i\omega_{n}-T_{A_{2}}}\label{eBii},
\end{equation}
which in combination with the expression for $Z_1(1/2)$
gives,
\begin{equation}
G_{12}^{O(1/N)}(i\omega_{n})=\frac{K_{1}(E_{01})}{i\omega_{n}-T_{A_{1}}}(-i\sqrt{\Gamma_1 \Gamma_2})\frac{K_{2}(E_{02})}{i\omega_{n}-T_{A_{2}}}.\label{eBiii}
\end{equation}
Note that the term $\sqrt{\Gamma_1\Gamma_2}$ in Eq. \ref{eBiii} results from the conduction
line in Fig. \ref{fig:M1}.


\begin{thebibliography}{10}

\bibitem{glazman-review} M. Pustilnik and L. Glazman, J. Phys. Condens. Matter 16, R513 (2004).

\bibitem{gold} D. Goldhaber-Gordon et al., Phys. Rev. Lett. {\bf 81}, 5225 (1998);
D. Goldhaber-Gordon et al., Nature {\bf 391}, 156 (1998).

\bibitem{cronenwett} S. Cronenwett et al., Science {\bf 540}, 281 (1998);
W. G. van der Wiel et al., Science {\bf 289}, 2105 (2000).

\bibitem {gold1} R. M. Potok, I. G. Rau, H. Shtrikman, Y. Oreg, D. Goldhaber-Gordon,
Nature {\bf 446}, 167 (2007).

\bibitem{gold_nano} C. H. L. Quay, John Cumings, Sara Gamble, R. de Picciotto, H. Kataura, D. Goldhaber-
Gordon, Phys. Rev. B 76, 245311 (2007).

\bibitem{chen-2004} J. C. Chen, A.M. Chang, and M. R. Melloch, Phys. Rev. Lett. {\bf 92}, 176801 (2004).

\bibitem{sigrist-2004} M. Sigrist, A. Fuhrer, T. Ihn, K. Ensslin, S. E. Ulloa, W.Wegscheider, and M. Bichler, Phys. Rev. Lett. {\bf 93}, 066802 (2004).

\bibitem{chen-review} A M Chang and J C Chen, Rep. Prog. Phys. {\bf 72}, 096501 (2009).

\bibitem{holleitner-2004} A. W. Holleitner, A. Chudnovskiy, D. Pfannkuche, K. Eberl, and R. H. Blick, Phys. Rev. B {\bf 70}, 075204 (2004).

\bibitem{hubel-2008} A. Hubel, K. Held, J. Weis, and K. v. Klitzing, Phys. Rev. Lett. {\bf 101}, 186804 (2008).

\bibitem{lee-2010} Minchul Lee, Mahn-Soo Choi, Rosa Lopez, Ramon Aguado,
Jan Martinek, and Rok Zitko, Phys. Rev. B {\bf 81}, 121311 (2010).

\bibitem{sela-2009} Eran Sela and Ian Affleck, Phys. Rev. Lett. {\bf 103}, 087204 (2009).

\bibitem{sela-2009qcp} Eran Sela and Ian Affleck, Phys. Rev. Lett. {\bf 102}, 047201 (2009).

\bibitem{Ireneusz-2010} Ireneusz Weymann, Phys. Rev. B {\bf 78}, 045310 (2008).

\bibitem{malecki-sela-affleck} Justin Malecki, Eran Sela, and Ian Affleck, arXiv:1009.0860 (2010).

\bibitem{oguri} Yunori Nisikawa and  Akira Oguri,J. Phys.: Conf. Ser.  {\bf 150},022066 (2009).

\bibitem{short} R. M. Konik, Phys. Rev. Lett. {\bf 99}, 076602 (2007).

\bibitem{long} R. M. Konik, New J. Phys. {\bf 9}, 257 (2007).


\bibitem{zitko} R. Zitko and J. Bonca, Phys. Rev. B {\bf 76}, 241305(R) (2007).

\bibitem{zitko-bonea-2006} Rok Zitko and Janez Bonca, Phys. Rev. B {\bf 74}, 045312 (2006).

\bibitem{hofstetter} Chung-Hou Chung and W. Hofstetter, Phys. Rev. B {\bf 76}, 045329 (2007).

\bibitem{ding} G.-H. Ding, Fei Ye, and B. Dong, J. Phys. Cond.-Matt. {\bf 21}, 455303 (2009).

\bibitem{ding-ye} Ding Guo-Hui and Ye Fei, Chin. Phys. Lett. {\bf 24}, 2926 (2007).

\bibitem{le-hur} R. Lopez, D. Sanchez, M. Lee, M.-S. Choi, P. Simon, and K. Le Hur, Phys. Rev. B 71, 115312 (2005).

\bibitem{vernek} E. Vernek, N. Sandler, S.E. Ulloa, and E.V. Anda, Physica E {\bf 34}, 608 (2006).

\bibitem{ding-kim} Guo-Hui Ding, Chul Koo Kim, and Kyun Nahm, Phys. Rev. B {\bf 71}, 205313 (2005).

\bibitem{luis-2006} Luis G. G.V. Dias da Silva, Nancy P. Sandler, Kevin Ingersent, and Sergio E. Ulloa, Phys. Rev. Lett. {\bf 97}, 096603 (2006).

\bibitem{luis-2008} Luis G. G. V. Dias da Silva, Kevin Ingersent, Nancy Sandler, and Sergio E. Ulloa1, Phys. Rev. B {\bf 78}, 153304 (2008).

\bibitem{sela-affleck-2009p} Eran Sela and Ian Affleck, Phys. Rev. B {\bf 79}, 125110 (2009).

\bibitem{komnik} J. P. Dahlhaus, S. Maier, and A. Komnik, Phys. Rev. B {\bf 81}, 075110 (2010).

\bibitem{kulk} M. Kulkarni and R. Konik, Phys. Rev. B {\bf 83}, 245121 (2011).

\bibitem{bickers} N.E. Bickers, Review of Modern Physics \textbf{59},
No.4, October 1987.

\bibitem{brandt} U. Brandt, H. Keiter, and Fu-Sui Liu, Z. Phys. B - Condensed Matter \textbf{58} , 267-281 (1985).

\bibitem{mahen} Gerald D. Mahan, Many Particle Physics, Plenum Press, NY (1981)

\bibitem{fetter} Alexander L. Fetter, John Dirk Walecka, Quantum theory of many-particle systems, Dover Publications, 2003

\bibitem{orland} J. W. Negele, Henri Orland, Quantum Many-Particle Systems, (1988)

\bibitem{mwprl} Y. Meir and N. S. Wingreen., Phys. Rev. Lett. {\bf 68}, 2512 (1992)

\bibitem{aditi} Z. Ratiani, A. Mitra, Phys. Rev. B. {\bf 79}, 245111 (2009)

\bibitem{pcoleman} A. Posazhennikova, B. Bayani, P. Coleman, Phys. Rev. B. {\bf 75}, 245329 (2007)

\bibitem{pcoleman1} E. Lebanon, P. Coleman, Phys. Rev. B. {\bf 76}, 085117 (2007)

\end{thebibliography}
\end{document}